\documentclass[submission,copyright]{eptcs}
\providecommand{\event}{Graphs as Models 2017} 
\providecommand{\dragan}{Bo{\v{s}}na{\v{c}}ki }

\usepackage{breakurl}             
\usepackage{underscore}           
\usepackage{color, soul}
\usepackage{graphicx}
\usepackage{float}
\usepackage{epstopdf}
\usepackage{xspace}
\usepackage{wrapfig}
\usepackage{todonotes}
\usepackage{float}
\usepackage{booktabs}
\usepackage[group-separator={,},group-minimum-digits=3]{siunitx}
\usepackage{tikz}
\usepackage{pgfplots}
\usepackage{mathtools}
\usetikzlibrary{arrows,automata,decorations.pathreplacing}

\DeclarePairedDelimiter\floor{\lfloor}{\rfloor}

\newcommand{\GPUexplore}{\textsc{GPUexplore}\xspace}

\newcommand{\GH}{\textsc{Gh}\xspace}
\newcommand{\GHcbs}{\textsc{Gh-cbs}\xspace}

\title{On the Scalability of the \textsc{GPUexplore}\\ Explicit-State Model Checker}
\author{Nathan Cassee \qquad\qquad Thomas Neele \qquad\qquad Anton Wijs\thanks{We gratefully acknowledge the support of NVIDIA Corporation with the donation of the GeForce Titan X used for this research.}
\institute{Eindhoven University of Technology\\ Eindhoven, The Netherlands}
\email{N.W.Cassee@student.tue.nl \{T.S.Neele, A.J.Wijs\}@tue.nl}
}
\def\titlerunning{On the Scalability of the \textsc{GPUexplore}\ Explicit-State Model Checker}
\def\authorrunning{N.W. Cassee, T.S. Neele \& A.J. Wijs}
\begin{document}
\maketitle

\begin{abstract}
The use of graphics processors (GPUs) is a promising approach to speed up model checking to such an extent that it becomes feasible to instantly verify software systems during development. \GPUexplore is an explicit-state model checker that runs all its computations on the GPU. Over the years it has been extended with various techniques, and the possibilities to further improve its performance have been continuously investigated. In this paper, we discuss how the hash table of the tool works, which is at the heart of its functionality. We propose an alteration of the hash table that in isolated experiments seems promising, and analyse its effect when integrated in the tool. Furthermore, we investigate the current scalability of \GPUexplore, by experimenting both with input models of varying sizes and running the tool on one of the latest GPUs of NVIDIA.
\end{abstract}

\section{Introduction}

Model checking~\cite{principlesofmodelchecking} is a technique to systematically determine whether a concurrent system adheres to desirable functional properties. There are numerous examples in which it has been successfully applied, however, the fact that it is computationally very demanding means that it is not yet a commonly used procedure in software engineering. Accelerating these computations with graphics processing units (GPUs) is one promising way to model check a system design in mere seconds or minutes as opposed to many hours.

\GPUexplore~\cite{GPUExplore,GPUexplore-safety,GPUExplore2} is a model checker that performs all its computations on a GPU. Initially, it consisted of a state space exploration engine~\cite{GPUExplore}, which was extended to perform on-the-fly deadlock and safety checking~\cite{GPUexplore-safety}.
Checking liveness properties has also been investigated~\cite{Wijs2016}, with positive results, but liveness checking has yet to be integrated in the official release of the tool. Finally, in order to reduce the memory requirements of \GPUexplore, partial order reduction has been successfully integrated~\cite{gpupor}.

Since the first results achieved with \GPUexplore~\cite{GPUExplore}, considerable progress has been made. For instance, the original version running on an NVIDIA K20 was able to explore the state space of the \texttt{peterson7} model in approximately 72 minutes. With many improvements to \GPUexplore's algorithms, reported in~\cite{GPUExplore2}, the GPU hardware and the CUDA compiler, this has been reduced to 16 seconds. With these levels of speed-up, it has become much more feasible to interactively check and debug large models. Furthermore, GPU developments continue and many options can still be investigated. 

Performance is very important for a tool such as \GPUexplore. However, so far, the scalability of the tool has not yet been thoroughly investigated. For instance, currently, we have access to a NVIDIA Titan X GPU, which is equipped with 12 GB global memory, but for all the models we have been using so far, 5 GB of global memory suffices as the models used for the run-time analysis of \GPUexplore do not require more than 5GB for a state space exploration. In the current paper, we report on results we have obtained when scaling up models to utilise up to 12 GB.

In addition, we also experimentally compared running \GPUexplore on a Titan X GPU with the Maxwell architecture, which was released on 2015, with \GPUexplore running on a Titan X GPU with the Pascal architecture, which was released a year later. This provides insights regarding the effect recent hardware developments have on the tool.

Finally, we analyse the scalability of a critical part of the tool, namely its \emph{hash table}. This structure is used during model checking to keep track of the progress made through the system state space. Even a small improvement of the hash table design may result in a drastic improvement in the performance of \GPUexplore. Recently, we identified, by conducting isolated experiments, that there is still potential for further improvement~\cite{GamPaperHashtables}. In the current paper, we particularly investigate whether changing the size of the so-called \emph{buckets}, i.e., data structures that act as containers in which states are stored, can have a positive effect on the running time.

The structure of the paper is as follows. In Section~\ref{sec:relatedwork}, we discuss related work. Next, an overview of the inner working of \GPUexplore is presented in Section~\ref{sec:gpus}. The hash table and its proposed alterations are discussed in Section~\ref{sec:hashtable}. After that, we present the results we obtained through experimentation in Section~\ref{sec:Experiments}. Finally, conclusions and pointers to future work are given in Section~\ref{sec:conclusion}.

\section{Related work}
\label{sec:relatedwork}

In the literature, several different designs for parallel hash tables can be found. First of all, there is the hash table for GPUs proposed by Alcantara \emph{et al.}~\cite{AlcChpt}, which is based on Cuckoo hashing. Secondly, Laarman \emph{et al.}~\cite{laarman-hashtable} designed a hash table for multi-core shared memory systems. Their implementation was later used as a basis for the hash table underlying the \textsc{LTSmin} model checker. Other lock-free hash tables for the GPU are those proposed by Moazeni \& Sarrafzadeh~\cite{linkedList}, by Bordawekar \cite{pres} and by Misra \& Chaudhuri~\cite{PerfEvlLockFree}. Cuckoo hashing as implemented by Alcantara \emph{et al.} is publicly available as part of the CUDPP library \footnote{\url{http://cudpp.github.io/}}. Unfortunately, to the best of our knowledge, there are no implementations available of the other hash table designs.

Besides \GPUexplore~\cite{GPUExplore2}, there are several other GPU model checking tools. Bartocci \emph{et al.}~\cite{Spin} developed an extension for the SPIN model checker that performs state-space exploration on the GPU. They achieved significant speed-ups for large models.

A GPU extension to the parallel model checking tool DIVINE, called DIVINE-CUDA, was developed by Barnat \emph{et al.}~\cite{divinegpu}. To speed-up the model checking process, they offload the cycle detection procedure to the GPU. Their tool can even benefit from the use of multiple GPUs. DIVINE-CUDA achieves a significant speed-up when model checking properties that are valid.

Edelkamp and Sulewski address the issues arising from the limited amount of global memory available on a GPU. In~\cite{EdSul10}, they implement a hybrid approach in a tool called \textsc{CuDMoC}, using the GPU for next state computation, while keeping the hash table in the main memory, to be accessed by multiple threads running on the Central Processing Unit (CPU). In~\cite{edelkamp-delayedduplicate}, they keep part of the state space in the global memory and store the rest on disk. The record on disk can be queried through a process they call \emph{delayed duplicate detection}. Even though disk access causes overhead, they manage to achieve a speed-up over single-threaded tools.

Edelkamp \emph{et al.}~\cite{Edelkamp,bosnacki.edelkamp.2010} also applied GPU programming to probabilistic model checking. They solve systems of linear equations on the GPU in order to accelerate the value iteration procedure. GPUs are well suited for this purpose, and can enable a speed-up of 18 times over a traditional CPU implementation.

Wu \emph{et al.}~\cite{singaporePaper} have developed a tool called GPURC that performs the full state-space exploration process on the GPU, similar to \GPUexplore. Their implementation applies \emph{dynamic parallelism}, a relatively new feature in CUDA that allows launching of new kernels from within a running kernel. Their tool shows a good speed-up compared to traditional single-threaded tools, although the added benefit of dynamic parallelism is minimal.

Finally, GPUs are also successfully applied to accelerate other computations related to model checking.
For instance, Wu \emph{et al.}~\cite{gpu-counterexamples} use the GPU to construct counter-examples,
and state space decomposition and minimisation are investigated in~\cite{wijs.tacas15,wijs.cav14,WijKaBo16}. For probabilistic model checking, {\v C}e{\v s}ka \emph{et al.}~\cite{Ceska2016} implemented GPU accelerated parameter synthesis for parameterized continous time Markov chains.

\section{GPUs and \GPUexplore}
\label{sec:gpus}

\begin{figure}[t]
	\centering
	\begin{minipage}{0.5\textwidth}
	\scalebox{0.8}{
		\begin{tikzpicture}[->,>=stealth',shorten >=0pt,auto,node distance=0.3cm,semithick]
			\tikzstyle{every state}=[draw,minimum size=15pt]
			
			\node at (0,2.5) {\scriptsize producer};
			\node at (2.75,2.5) {\scriptsize consumers};
			\node[state]    (s0)     at (0,1.5)     {$p_0$};
			\node[state]    (s1)     at (0,0)       {$p_1$};
			\node[state]    (c00)    at (2,0)     {$c_0$};
			\node[state]    (c01)    at (3.5,0)       {$c_1$};
			\node[state]    (c10)    at (2,1.5)   {$c_0$};
			\node[state]    (c11)    at (3.5,1.5)     {$c_1$};
			
			\path   node [left=of s0]  {} edge (s0)
					node [left=of c00] {} edge (c00)
					node [left=of c10] {} edge (c10)  
					(s0)    edge[bend right,swap]    node {gen\_work}  (s1)
					(s1)    edge[bend right,swap]    node {send}       (s0)
					(c00)   edge[bend left]          node {rec}        (c01)
					(c01)   edge[loop right]         node {work}       (c01)
					(c01)   edge[bend left]          node {$\tau$}     (c00)
					(c10)   edge[bend left]          node {rec}        (c11)
					(c11)   edge[loop right]         node {work}       (c11)
					(c11)   edge[bend left]          node {$\tau$}     (c10);
		\end{tikzpicture}
	}
	\end{minipage}
	\begin{minipage}{0.4\textwidth}
		\scriptsize
		\begin{verbatim}
			par using
			    send * rec *  _  -> trans,
			    send *  _  * rec -> trans
			in
			    "producer.aut"
			    ||
			    "consumer.aut"
			    ||
			    "consumer.aut"
			end par
		\end{verbatim}
	\end{minipage}
	\caption{Example of LTS network with one producer and two consumers. On the right, the communication between the LTSs is specified using the EXP syntax~\cite{cadp}. Here, \texttt{producer.aut} and \texttt{consumer.aut} are files containing the specification of the producer and the consumer respectively.}
	\label{fig:ltsnetwork}
\end{figure}
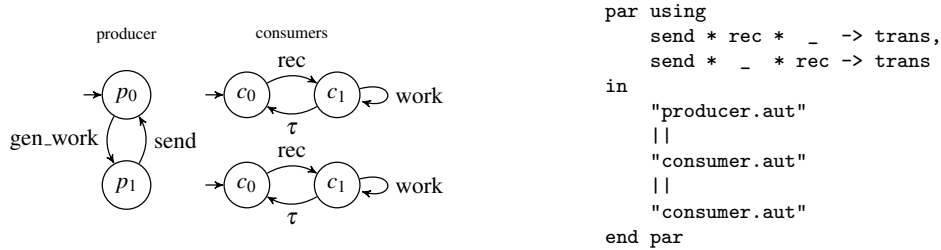

\GPUexplore~\cite{GPUExplore,GPUexplore-safety,GPUExplore2} is an explicit-state model checker that practically runs entirely on a GPU (only the general progress is checked on the host side, i.e.\ by
a thread running on the CPU). It is written in CUDA C, an extension of C offered by \textsc{Nvidia}.
CUDA (Compute Unified Device Architecture) provides an interface to write applications for \textsc{Nvidia}'s GPUs.
\GPUexplore takes a \emph{network of Labelled Transition Systems} (LTSs)~\cite{Lang06} as input, and can construct the synchronous product of those LTSs using many threads in a Breadth-First-Search-based exploration, while optionally checking on-the-fly for the presence of deadlocks and violations of safety properties. A (negation of a) safety property can be added as an automaton to the network.

An LTS is a directed graph in which the nodes represent states and the edges are transitions between the states. Each transition has an action label representing an event leading from one state to another. An example network is shown in Figure~\ref{fig:ltsnetwork}, where the initial states are indicated by detached incoming arrows. One producer generates work and sends it to one of two consumers. This happens by means of synchronisation of the `send' and `rec' actions. The other actions can be executed independently. How the process LTSs should be combined using the relevant synchronisation rules is defined on the right in Figure~\ref{fig:ltsnetwork}, using the syntax of the \textsc{Exp.open} tool~\cite{Lang06}. The state space of this network consists of 8 states and 24 transitions.

The general approach of \GPUexplore to perform state space exploration is discussed in this section, leaving out many of the details that are not relevant for understanding the current work. The interested reader is
referred to~\cite{GPUExplore,GPUexplore-safety,GPUExplore2}.

\begin{figure}[t]
	\centering
	\scalebox{0.85}{
	\includegraphics[width=0.5\textwidth]{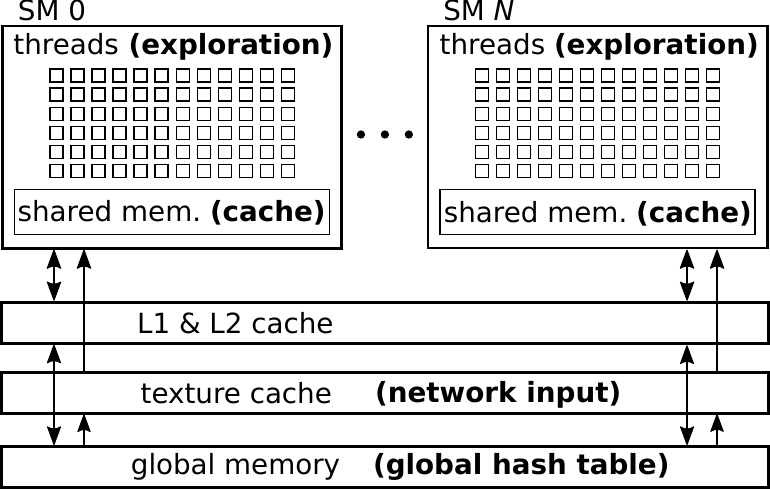}
	}
	\caption{Schematic overview of the GPU hardware architecture and \GPUexplore}
	\label{fig:gpuarchitecture}
\end{figure}

In a CUDA program, the host launches CUDA functions called \emph{kernels}, that are to be executed many times in parallel by a specified number of GPU threads. Usually, all threads run the same kernel using different parts of the input data, although some GPUs allow multiple different kernels to be executed simultaneously (\GPUexplore does not use this
feature). Each thread is executed by a streaming processor (SP). Threads are grouped in \emph{blocks} of a predefined size. Each block is 
assigned to a streaming multiprocessor (SM). An SM consists of a fixed number of SPs (see Figure~\ref{fig:gpuarchitecture}).

Each thread has a number of
on-chip registers that allow fast access. The threads in a block together share memory to exchange data, which is located in the (on-chip) \emph{shared memory} of an SM.
Finally, the blocks can share data using the \emph{global memory} of the GPU, which is relatively large, but slow, since it is off-chip.
The global memory is used to exchange data between the host and the kernel.
The \textsc{GTX Titan X}, which we used for our experiments, has 12 GB global memory and 24 SMs, each having 128 SPs (3,072 SPs in total).

Writing well-performing GPU applications is challenging, due to the execution model of GPUs, which is \emph{Single Instruction Multiple Threads}.
Threads are partitioned in groups of 32 called \emph{warps}. The threads in a warp run in lock-step, sharing a program counter, so they always execute the same program instruction.
Hence, thread divergence, i.e.\ the phenomenon of threads being forced to execute different instructions (e.g., due to if-then-else constructions) or to access physically distant
parts of the global memory, negatively affects performance.

Model checking tends to introduce divergences frequently, as it requires combining the behaviour of the processes in the network, and accessing and storing state vectors of the system state
space in the global memory. In \GPUexplore, this is mitigated by combining relevant network information as much as possible in 32-bit integers, and storing these as textures, that only allow
read access and use a dedicated cache to speed up random accesses. \par 
Furthermore, in the global memory, a hash table is used to store state vectors (Figure~\ref{fig:gpuarchitecture}). The currently used hash table has been designed to
optimise accesses of entire warps: the space is partitioned into buckets consisting of 32 integers, precisely enough for one warp to fetch a bucket with one combined memory access. State vectors
are hashed to buckets, and placed within a bucket in an available slot. If the bucket is full, another hash function is used to find a new bucket.
Each block accesses the global hash table to collect vectors that still require exploration.
\par To each state vector with $n$ process states, a \emph{group} of $n$ threads is assigned to construct its
successors using fine-grained parallelism.
Since access to the global memory is slow, each block uses a dedicated state cache (Figure~\ref{fig:gpuarchitecture}). It serves to store and collect newly produced state vectors, that are subsequently moved to the global hash table in batches. With the cache, block-local duplicates can be detected.

\section{The \GPUexplore Hash Table}
\label{sec:hashtable}

States discovered during the search exploration phase of \GPUexplore are inserted into a global memory hash table. This hash table is used to keep track of the open and closed sets maintained during the breadth first search based exploration of the state space. \par 

Since many accesses (reads and writes) to the hash table are performed during state-space exploration, its performance is critical for our model checker. In order to allow for efficient parallel access, the hash table should be lock-free. To prevent corruption of state vectors, insertion should be an atomic operation, even when a state vector spans multiple 32 bit integers.

Given these requirements, we have considered several lock-free hash table implementations. One of them, proposed by Alcantara \emph{et al.}~\cite{AlcChpt}, uses so-called \emph{Cuckoo hashing}. \par 
With Cuckoo hashing a key is hashed to a bucket in the hash table, and in case of a collision the key that is already in the bucket is evicted and rehashed using another hash function to a different bucket. Re-insertions continue until the last evicted key is hashed to an empty bucket, until all hash functions are exhausted or until the chain of re-insertions becomes too long~\cite{cuckoo}. \par 
The other hash table we considered is the one originally designed for \GPUexplore~\cite{GPUExplore2}; we refer to its hashing mechanism as \GPUexplore hashing. We experimentally compared these two hash tables, and from these comparisons we concluded that while Cuckoo hashing on average performs better, it does not meet all the demands needed by \GPUexplore. However, based on the performance evaluation a possible performance increase has been identified for \GPUexplore hashing~\cite{GamPaperHashtables}. This section discusses the proposed modification, and its implementation.

\subsection{\GPUexplore Hashing }

\begin{figure}[H]
\centering
\includegraphics[width=.8\textwidth]{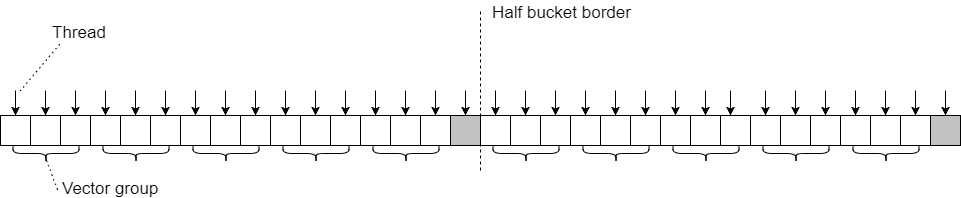}
\caption{Division of threads in a warp over a bucket of size 32 in \GPUexplore 2.0}
\label{fig:static32}
\end{figure}

The \GPUexplore hash table consists of two parts: the storage used to store the discovered vectors and the set of hash constants used by the hash function to determine the position of a vector in the hash table. The memory used to store vectors is divided into buckets with a size of 32 integers. Each bucket is additionally split into two equally sized half buckets. Therefore a single bucket can store $ 2 * \floor*{\frac{16}{{\tt vector\_length}}}$ vectors.
The reason for writing vectors to buckets with half-warps (a group of 16 threads) is that in many cases, atomic writes of half-warps are scheduled in an uninterrupted sequence~\cite{GPUExplore2}. This results in vectors consisting of multiple integers to be written without other write operations corrupting them.
It should be noted that the \GPUexplore hash table uses \emph{closed hashing}, i.e., the vectors themselves are stored in the hash table, as opposed to pointers to those vectors.
\par 

When inserting, a warp of 32 threads inserts a single vector into a bucket. The way threads are divided over a bucket can be observed in Figure \ref{fig:static32}, this figure visualizes a single bucket of the \GPUexplore hash table for a vector length of 3. Each thread in a warp is assigned to one integer of the bucket, and to one integer of the vector. This assignment is done from left to right per half bucket. For this example the first 3 threads, i.e., the first vector group, is assigned to the first slot in the bucket, and the first thread in a vector group is assigned to the first integer of the vector. By assigning every thread to a single part of the vector and of the bucket each thread has a relatively simple task, which can be executed in parallel. \par

The insertion algorithm first hashes the vector to insert to determine the bucket belonging to the vector. Each thread checks its place in the bucket and compares the integer on that position to the corresponding integer of the vector. After comparing, the insertion algorithm then uses CUDA warp instructions to quickly exchange data between all 32 threads in the warp to determine whether a single vector group of threads has found the vector. If the vector has been found the insertion algorithm terminates, if the vector has not been found the algorithm continues. \par 

\begin{figure}[H]
\centering
\includegraphics[width=.8\textwidth]{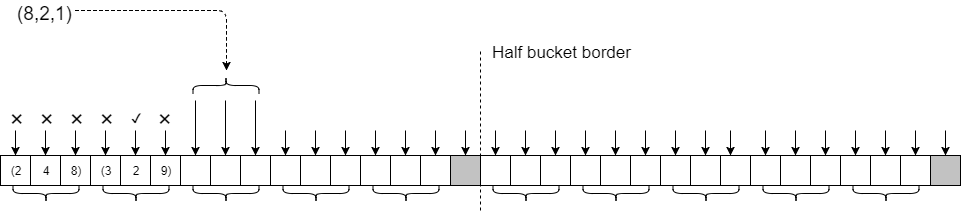}
\caption{Example of inserting the vector $(8, 2, 1)$ into the  \GPUexplore 2.0 hash table.}
\label{fig:static32-ex}
\end{figure}

Figure \ref{fig:static32-ex} shows an example of the vector $(8, 2, 1)$ being inserted into the \GPUexplore hash table where the first two slots are already occupied. Because the first two slots are already occupied the first six threads compare their element of the vector to their element of the bucket. The icons above the arrows indicate the result of these comparisons. As can be seen there is one thread that has a match, however, because not all elements in the slot match the insertion algorithm does not report a find. \par 

If the vector is not found in the bucket, the insertion algorithm selects the first free slot in the bucket, in this case the third slot. This selection procedure can be done efficiently using CUDA warp instructions. Next, the associated threads attempt to insert the vector $(8, 2, 1)$ into the selected slot using a compare and swap operation. If the insertion fails because another warp had claimed that slot already for another vector, the algorithm takes the next free slot in the bucket. \par 

If a bucket has no more free slots the next set of hash constants is used, and the next bucket is probed. This is repeated until all hash constants have been used, and if no insertion succeeds into any of the buckets, the insertion algorithm reports a failure and the exploration stops. \par 

In \GPUexplore 2.0 the hash table is initialized using eight hash functions. After each exploration step, blocks that have found new vectors use the insertion algorithm to insert any new vectors they found into the global hash table. The hash table is therefore a vital part of \GPUexplore. \par 

Buckets with a length of 32 integers have been chosen because of the fact that warps in CUDA consist of 32 threads. This way, every integer in a bucket can be assigned to a single thread. Besides, this design choice also allows for coalesced memory access: when a warp accesses a continuous block of 32 integers, this operation can be executed in a single memory fetch operation~\cite{cuda}. Uncoalesced accesses, on the other hand, have to be done individually after each other. By coalescing the accesses, the available bandwidth to global memory is used efficiently.

\subsection{Configurable Bucket Size}

While the current implementation of the hash table makes it possible for \GPUexplore to achieve a considerable increase in performance over CPU-based explicit-state model checkers~\cite{GPUExplore2}, it suffers from one disadvantage.
Namely, after initially scanning the bucket, only $x$ threads, where $x$ is the vector length, are active at a time. The other $32 - x$ threads are inactive while they await the result of the atomic insertion of the active group. If the insertion fails but there is still a free slot in the hash table, another group of $x$ threads becomes active to attempt an atomic insertion, while the remaining $32 - x$ threads again await the result of this operation. \par

\begin{figure}[H]
\centering
\includegraphics[width=.9\textwidth]{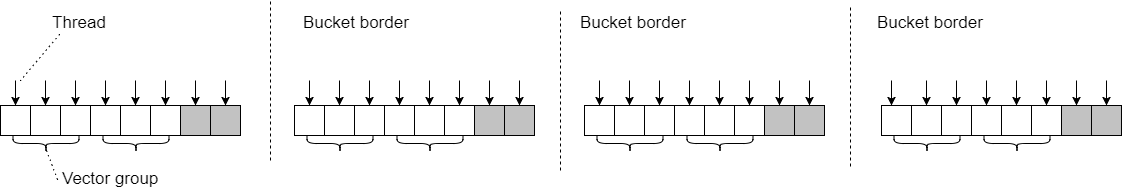}
\caption{Division of threads in a warp over buckets of size 8}
\label{fig:dynamic-buckets}
\end{figure}

Therefore, for buckets with a lot of free slots where insertions fail the majority of threads in the warp are inactive. Furthermore, the vector size generally does not exceed four integers, which means that when attempting to atomically insert a vector, the majority of threads in a warp is inactive. Ergo, one possible improvement over the \GPUexplore hash table is to reduce the bucket size, so that there are fewer slots per bucket, and therefore, fewer threads are needed to insert a single vector. As a single insertion still uses one thread per integer in the bucket, in turn more vectors can be inserted in parallel.\par 

Figure \ref{fig:dynamic-buckets} shows what the division of threads over a warp looks like if buckets of size 8 instead of 32 are used. As can be observed, a warp can insert four elements in parallel, as in this diagram each group of 8 threads inserts a different vector and accesses different buckets in global memory. \par

The logical consequence of this improvement is that after scanning a bucket, fewer threads are inactive while the vector is being inserted into a free slot. If we suppose that the vector size for a certain model is 3, and that the new bucket size is 8, then while inserting a vector using the \GPUexplore hash table $32 - 3 = 29$ threads are inactive. On the other hand, if four buckets of size 8 are simultaneously accessed, only $32 - 3\cdot 4 = 20$ threads are inactive, and four vectors are simultaneously processed, as opposed to only one.\par 

However, while more threads can be active at the same time, smaller buckets also lead to thread divergence within a warp. First of all, of course, accessing different buckets simultaneously likely leads to uncoalesced memory accesses. Furthermore, it is also possible that in an insertion procedure, one group needs to do more work than another in the same warp. For instance, consider that the first group in the warp fails to find its vector in the designated bucket, and also cannot write it to the bucket since the latter is full. In that case, the group needs to fetch another bucket. At the same time, another group in the warp may find its vector in the designated bucket, and therefore be able to stop the insertion procedure. In such a situation, these two groups will diverge, and the second group will have to wait until the first group has finished inserting. This means that the use of smaller buckets can only be advantageous if the performance increase of the smaller buckets outweighs the performance penalty of divergence. In this paper, we address whether this is true or not in practical explorations. \par

The suggested performance increase has been experimentally validated by comparing an implementation of the hash table with varying bucket size to the original \GPUexplore 2.0 hash table, both in isolation and as a part of \GPUexplore. The results of this comparison are presented and discussed in section \ref{sec:Experiments}.

\section{Experiments}
\label{sec:Experiments}
In this section we discuss our experiments to evaluate the scalability of \GPUexplore. In Section~\ref{sec:expbucket}, we report on experiments to evaluate the effect of the bucket size on the runtimes of \GPUexplore. In the next two sections, our goal is to determine whether \GPUexplore scales well when varying the size of the model and the performance of the hardware, respectively. For most of the experiments, we use an NVIDIA Titan X (Maxwell) installed on a machine running \textsc{Linux Mint 17.2}. The number of blocks is set to 6,144, with 512 threads per block. The hash table is allocated 5GB in global memory. 

For our benchmarks, we selected a varied set of models from the CADP toolset~\cite{cadp}, the mCRL2 toolset~\cite{mcrl2} and the BEEM database~\cite{beem}. The models with a .1 suffix have been modified to obtain a larger state space.

\subsection{Varying the Bucket Size}
\label{sec:expbucket}

To test different bucket sizes two types of experiments have been performed. First, the hash table with varying bucket sizes has been tested in isolation, where the time taken to insert 100,000,000 elements has been measured. Second, \GPUexplore has been modified such that it uses a hash table with modifiable bucket size. This bucket size can be set at compile time. The performance of this version of \GPUexplore has been compared to the original \GPUexplore w.r.t.\ the exploration of several input models. \par

\begin{figure}
	\centering
	\begin{tikzpicture}

	\begin{axis}[
	ylabel=Runtime (ms),
	xlabel=Duplication in sequence,
	tick label style={font=\footnotesize},
	xtick={0,10, ..., 100},
	xmin=0, xmax=100,
	ymin=0,
	tick align=outside,
	xtick pos=left,
	width=0.6\textwidth,
	height=8cm,
	mark size=0.6pt,
	cycle list={%
		red,mark=*\\%
		cyan,mark=diamond*\\%
	}
	]
	\pgfplotsinvokeforeach{gpuexplore,gpuexplorebucketed}{
		\addplot table[x=division,y=#1] {hashtable.csv};
	}
	
	\legend{\GH, \GHcbs}
	\end{axis}
	\end{tikzpicture}
	\caption{Results of inserting a sequence of 100,000,000 randomly generated integers into the hash tables of \GPUexplore (\GH) and \GPUexplore with configurable buckets (\GHcbs). For the bucketed version a bucket size of four integers has been used.}
	\label{fig:buckvsgputable}
\end{figure}
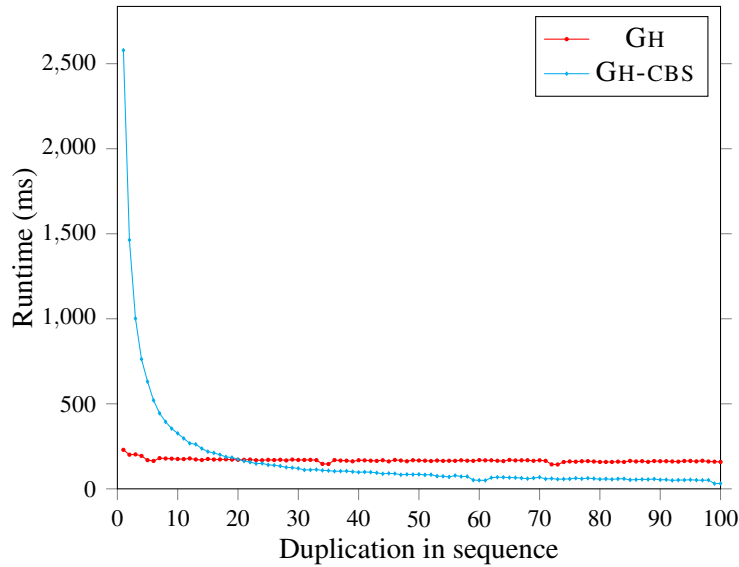

The input data for the performance evaluation of the hash tables in isolation is a set of sequences of randomly generated integers, each sequence consisting of 100,000,000 vectors with a length of 1 integer. The sequences vary in how often an element occurs in the sequence. A duplication of 1 means that every unique element occurs once in the sequence, and a duplication of 100 means that each unique element in the sequence occurs 100 times. Therefore a sequence with a duplication of 100 has $\frac{100,000,000}{100}$ unique elements. Note that this experiment tries to replicate state space exploration, where many duplicate insertions are performed when the fan-in is high, i.e., many transitions in the state space lead to the same state. So this experiment replicates the performance of the hash table for models that vary in the average fan-in of the states. For each sequence, we measured the total time required to insert all elements. \par 

The results of this comparison are depicted in Figure \ref{fig:buckvsgputable}. We refer to the standard \GPUexplore 2.0 hash table as \GH and the hash table with configurable bucket size as \GHcbs. In the experiment, \GHcbs with a bucket size of four integers has been compared to \GH. \GHcbs is slower for sequences where each element only occurs a few times. However, for sequences with a higher duplication degree \GHcbs starts to outperform \GH. After all, for the sequences with more duplication, less time is spent on atomically inserting elements, since most elements are already in the hash table. \GHcbs performs up to three times better than \GH. We will see later why the amount of duplication is relevant for model checking. \par

In addition to testing the performance of the hash tables in isolation, the performance of \GHcbs has been compared with standard \GPUexplore hashing as a part of \GPUexplore 2.0. The hash table underlying \GPUexplore, as implemented by Wijs, Neele and \dragan~\cite{GPUExplore2}, has been modified to allow configuration of the bucket size at compile time. We compared the time required for state space exploration of our implementation, \GPUexplore with configurable bucket size, with the original implementation of \GPUexplore 2.0~\cite{GPUExplore2}. \par 

Four bucket sizes have been compared to \GPUexplore 2.0, namely bucket sizes 4, 8, 16 and 32 integers. The relative performance with these bucket sizes has been recorded with the performance of \GPUexplore 2.0 as baseline. For each model the experiment has been run five times and the average running time of these five runs has been taken. \par 

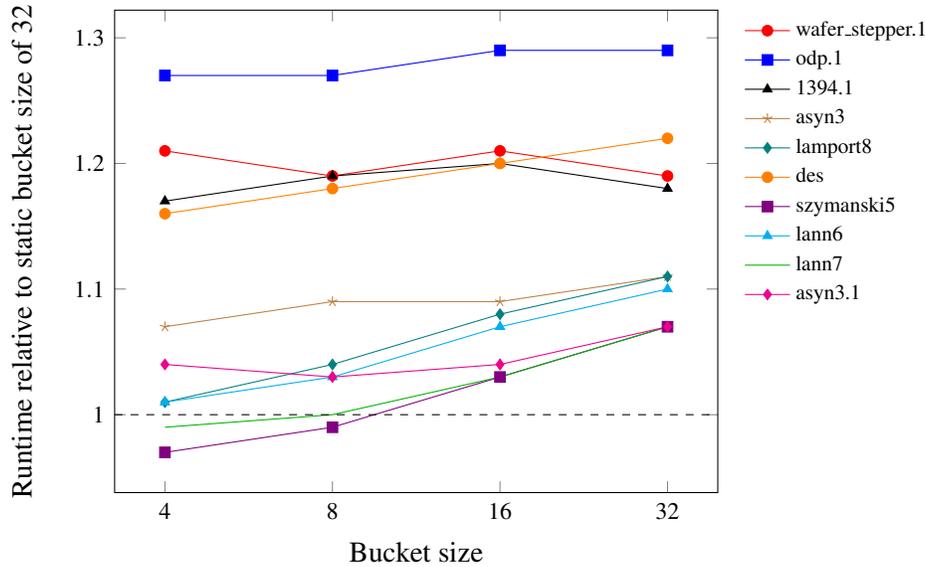
\begin{figure}
	\centering
	\begin{tikzpicture}
	\begin{semilogxaxis}[
	ylabel=Runtime relative to static bucket size of 32,
	xlabel=Bucket size,
	tick label style={font=\footnotesize},
	log ticks with fixed point,
	xtick=data,
	width=0.6\textwidth,
	height=8cm,
	legend style={draw=none,fill=none,at={(1,1)},legend cell align=left,font=\scriptsize},
	legend pos=outer north east,
	cycle list={%
		red,mark=*\\%
		blue,mark=square*\\%
		black,mark=triangle*\\%
		brown,mark=star\\%
		teal,mark=diamond*\\%
		orange,mark=*\\%
		violet,mark=square*\\%
		cyan,mark=triangle*\\%
		green!70!black,mark=start\\%
		magenta,mark=diamond*\\ %
	},
	]
	\pgfplotsinvokeforeach{waferstepper.1, odp.1, 1394.1, asyn3, lamport8, des, szymanski5,  lann6, lann7, asyn3.1}{
		\addplot table[x=bucket,y=#1] {buckets.csv};
	}
	\draw[black,dashed] ({rel axis cs:0,0}|-{axis cs:4,1}) -- ({rel axis cs:1,0}|-{axis cs:32,1});
	\legend{wafer\_stepper.1, odp.1, 1394.1, asyn3, lamport8, des, szymanski5, lann6, lann7, asyn3.1}
	\end{semilogxaxis}
	\end{tikzpicture}
	\caption{Relative runtime of \GPUexplore 2.0 with variable bucket size. The runtime \GPUexplore 2.0 with a fixed bucket size of 32 integers is used as a reference and is normalized to 1.}
	\label{fig:bucketsize}
\end{figure}

The result of these comparisons is illustrated in Figure \ref{fig:bucketsize}. As can be observed the total exploration time of \GPUexplore with configurable bucket size is for most models larger than the runtime of \GPUexplore 2.0. Only \emph{szymanski5} and \emph{lann7} show a small performance increase for bucket size 4. For the other instances, however, the new hash table causes a slow down of up to 30\%.\par

There are three reasons why the promising performance shown in the previous experiments is not reflected here. First, the increased complexity of the hash table negatively affects \emph{register pressure}, i.e., the number of registers each thread requires to execute the kernel. When the register usage in a kernel increases, the compiler may temporarily place register contents in global memory. This effect is not observed when the hash table is tested in isolation, as in that case, far fewer registers per thread are required. The increase in register pressure is also the reason that \GPUexplore with a configurable bucket size set to 32 is slower than \GPUexplore with a static bucket size of 32.\par

Furthermore, smaller bucket sizes result in more thread divergence and more uncoalesced memory accesses when reading from the hash table. Therefore, the available memory bandwidth is used less efficiently, leading to a drop in performance. Apparently, the increased potential for parallel insertions of vectors cannot overcome this drawback.\par

Lastly, while exploring the state-space, \GPUexplore only discovers duplicates if those states have several incoming transitions. On average, the models used for the experiments have a fan-in of 4 to 6, with some exceptions that have a higher fan-in of around 8 to 11. However, from Figure \ref{fig:buckvsgputable} it can be concluded that the hash table in isolation only starts to outperform the static hash table when each element is duplicated 21 times. This partly explains the performance seen in Figure \ref{fig:bucketsize}. \par

\subsection{Varying the Model Size}
In addition to experimentally comparing the effect of different bucket sizes, we also investigated how \GPUexplore 2.0 behaves when exploring state spaces of different size. We performed this experiment with two different models. The first is a version of the Gas Station model~\cite{gasstation} where the number of pumps is fixed to two. We varied the number of customers between two and twelve. None of these instances requires a state vector longer than two 32-bit integers.

The other model is a simple implementation of a ring-structured network, where one token is continuously passed forward between the nodes. Each node has two transitions to communicate with its neighbours and a further three internal transitions. Here, we varied the number of nodes in the network.

\begin{table}[t]
	\caption{Performance of \GPUexplore for the Gas Station and the Token Ring model while varying the amount of processes.}
	\label{tab:varying-size}
	\centering
	\begin{tabular}{lrrr}
		\toprule
		        \multicolumn{4}{c}{Gas Station}         \\
		\midrule
		N  &          states & time (s) &    states/sec \\
		\midrule
		2  & \num{      165} &    0.016 & \num{  10294} \\
		3  & \num{     1197} &    0.023 & \num{  51004} \\
		4  & \num{     7209} &    0.035 & \num{ 206812} \\
		5  & \num{    38313} &    0.062 & \num{ 621989} \\
		6  & \num{   186381} &    0.209 & \num{ 892039} \\
		7  & \num{   849285} &    0.718 & \num{1183246} \\
		8  & \num{  3680721} &    1.235 & \num{2981535} \\
		9  & \num{ 15333057} &    3.093 & \num{4957437} \\
		10 & \num{ 61863669} &   11.229 & \num{5509307} \\
		11 & \num{243104733} &   44.534 & \num{5458810} \\
		12 & \num{934450425} &  178.817 & \num{5225726} \\
		\bottomrule
	\end{tabular}%
	\hspace{0.4cm}
	\begin{tabular}{lrrr}
		\toprule
		        \multicolumn{4}{c}{Token Ring}          \\
		\midrule
		N  &          states & time (s) &    states/sec \\
		\midrule
		2  & \num{       12} &    0.027 & \num{    449} \\
		3  & \num{       54} &    0.047 & \num{   1138} \\
		4  & \num{      216} &    0.067 & \num{   3212} \\
		5  & \num{      810} &    0.086 & \num{   9402} \\
		6  & \num{     2916} &    0.113 & \num{  25702} \\
		7  & \num{    10206} &    0.180 & \num{  56571} \\
		8  & \num{    34992} &    0.275 & \num{ 127142} \\
		9  & \num{   118098} &    0.488 & \num{ 242225} \\
		10 & \num{   393660} &    1.087 & \num{ 362294} \\
		11 & \num{  1299078} &    6.394 & \num{ 203159} \\
		12 & \num{  4251528} &    8.345 & \num{ 509462} \\
		13 & \num{ 13817466} &    8.138 & \num{1697864} \\
		14 & \num{ 44641044} &   22.060 & \num{2023649} \\
		15 & \num{143489070} &   68.233 & \num{2102934} \\
		16 & \num{459165024} &  215.889 & \num{2126853} \\
		\bottomrule
	\end{tabular}
\end{table}

We executed the tool five times on each instance and computed the average runtime. The results are listed in Table~\ref{tab:varying-size}. For the smallest instances, the performance of \GPUexplore (measured in states/sec) is a lot worse compared to the larger instances. This has two main reasons. First of all, the relative overhead suffered from initialization and work scanning is higher. Second, the parallelism offered by the GPU cannot be fully exploited, because the size of one search layer is too small to occupy all blocks with work.

For the gas station model, peak performance is achieved for the instance with ten customers, which has 60 million states. For larger instances, the performance decreases slightly due to the increasing occupancy of the hash table. This leads to more hash collisions, therefore more time is lost on rehashing.

The results of the token ring model show another interesting scalability aspect. There is a performance drop between the instances with 10 nodes and 11 nodes. This is caused by the fact that the instance with 11 nodes is the smallest for which the state vector exceeds 32 bits in length. Longer state vectors lead to more memory accesses throughout the state-space generation algorithm.

\subsection{Speed-up Due to the Pascal Architecture}
The Titan X we used for most of the benchmarks is based on the Maxwell architecture, and was launched in 2015. Since then, NVIDIA has released several other high-end GPUs. Most aspects have been improved: the architecture has been revised, there are more CUDA cores on the GPU and there is more global memory available. To investigate how well \GPUexplore scales with faster hardware, we performed several experiments with a Titan X with the Maxwell architecture (TXM) and with a Titan X with the Pascal architecture (TXP). The latter was released in 2016, and the one we used is installed in the \textsc{Das-5} cluster~\cite{DAS5} on a node running \textsc{CentOS Linux 7.2}.

The TXM experiments were performed with 6,144 blocks, while for the TXP, \GPUexplore was set to use 14,336 blocks. The improvements of the hardware allows for \GPUexplore to launch more blocks. To evaluate the speed-ups compared to a single-core CPU approach, we also conducted experiments with the \textsc{Generator} tool of the latest version (2017-i) of the \textsc{Cadp} toolbox~\cite{cadp}. These have been performed on nodes of the \textsc{Das-5} cluster, which are equipped with an \textsc{Intel Haswell} E5-2630-v3 2.4 GHz CPU, 64 GB memory, and \textsc{CentOS Linux 7.2}.

The results are listed in Table \ref{tab:maxwell}. The reported runtimes are averages after having run the corresponding tool ten times. For most of the larger models, we see a speed-up of about 2 times when running \GPUexplore on a TXP compared to running it on a TXM. The average speed-up is 1.73. This indicates that \GPUexplore scales well with a higher memory bandwidth and a larger amount of CUDA cores.

\begin{table}[t]
	\caption{Performance comparison of single-core \textsc{Generator} of \textsc{Cadp} (\textsc{Gen}) and \GPUexplore running on a Titan X with the Maxwell architecture (TXM) and with the Pascal architecture (TXP).}
	\label{tab:maxwell}
	\centering
	\begin{tabular}{lrrrrrr}
		\toprule
		~                & ~               & \multicolumn{3}{c}{runtime (seconds)}          &    \multicolumn{2}{c}{speed-ups}      \\
		\cmidrule(lr){3-5} \cmidrule(lr){6-7}
		model            &          states & \textsc{Gen} & TXM & TXP & \textsc{Gen}-TXP & TXM-TXP  \\
		\midrule
		acs              & \num{     4764} &  4.17 &           0.05 &             0.06 & 71.91 &   0.89 \\
		odp              & \num{    91394} & 3.26 &            0.08 &             0.05 & 70.76 &   1.64 \\
		1394             & \num{   198692} & 2.81 &            0.20 &             0.15 & 18.95 &   1.36 \\
		acs.1            & \num{   200317} &  5.30  &          0.18 &             0.14 &  37.88 &  1.27 \\
		transit          & \num{  3763192} &  34.36   &         0.77 &             0.48 & 70.99  &  1.59 \\
		wafer\_stepper.1 & \num{  3772753} & 22.95 &            1.01 &             0.51 & 45.17 &   2.00 \\
		odp.1            & \num{  7699456} &   65.50   &        1.34 &             0.66 & 99.54 &   2.03 \\
		1394.1           & \num{ 10138812} &  82.71   &         1.42 &             0.83 & 99.66 &   1.71 \\
		asyn3            & \num{ 15688570} &  358.58    &        3.15 &             1.98 & 181.47 &   1.59 \\
		lamport8         & \num{ 62669317} & 1048.13    &         5.81 &             3.11 & 336.80 &   1.87 \\
		des              & \num{ 64498297} &  477.43   &        12.34 &             6.65 & 71.84 &   1.86 \\
		szymanski5       & \num{ 79518740} & 1516.71  &            7.48 &             3.90 & 389.10 &   1.92 \\
		peterson7        & \num{142471098} & 3741.87   &         31.60 &            15.74 & 237.81 &    2.01 \\
		lann6            & \num{144151629} &  2751.15   &        10.57 &             5.39 & 510.80 &   1.96 \\
		lann7            & \num{160025986} & 3396.19    &        16.67 &             8.41 & 403.92 &   1.98 \\
		asyn3.1          & \num{190208728} &  4546.84   &        31.03 &            15.37 & 295.92 &   2.02 \\
		\midrule
		~                &               ~ &                 ~ &   &       average & 183.91  &  1.73 \\
		\bottomrule
	\end{tabular}
\end{table}

Comparing \GPUexplore on a TXP with single-core CPU analysis, the average speed-up is 183.91, and if we only take state spaces into account consisting of at least 10 million states, the average speed-up is 280.81.
Considering that with multi-core model checking, linear speed-ups can be achieved~\cite{laarmanthesis}, this roughly amounts to using 180 and 280 CPU cores, respectively. This, in combination with the observation that frequently, speed-ups over 300 times and once even over 500 times are achieved, clearly demonstrates the effectiveness of using GPUs for explicit-state model checking.


\section{Conclusion and Future Work}
\label{sec:conclusion}

In this paper, we have reported on a number of scalability experiments we conducted with the GPU explicit-state model checker \GPUexplore. In earlier work, we identified potential to further improve its hash table~\cite{GamPaperHashtables}. However, experiments in which we varied the bucket size in \GPUexplore provided the insight that only for very specific input models, and only if the bucket size is set very small (4), some speed-up becomes noticeable. In the context of the entire computation of \GPUexplore, the additional register use per thread and the introduced uncoalesced memory accesses and thread divergence make it not worthwile to make the bucket size configurable. This may be different for other applications, as our experiments with the hash table in isolation point out that hashing can be made more efficient in this way.

Besides this, we have also conducted experiments with models of different sizes. We scaled up a Gas Station model and a Token Ring model and obtained very encouraging results; for the second model, \GPUexplore can generate up to 2.1 million states per second, and for the first model, at its peak, \GPUexplore is able to generate about 5.5 million states per second, exploring a state space of 934.5 million states in under three minutes. We believe these are very impressive numbers that demonstrate the potential of GPU model checking.

Finally, we reported on some experiments we conducted with new GPU hardware. The Titan X with the Pascal architecture from 2016 provides for our benchmark set of models an average speed-up of 1.73 w.r.t.\ the Titan X with the Maxwell architecture from 2015. We also compared the runtimes of \GPUexplore running on the Pascal Titan X with the CPU single-core \textsc{Generator} tool of the \textsc{Cadp} toolbox, and measured an average speed-up of 183.91 for the entire benchmark set of models, and of 280.81 for the models yielding a state space of at least 10 million states.
Often speed-ups over 300 times have been observed, and in one case even over 500 times.

\paragraph{Future work}
For future work, we will consider various possible extensions to the tool. First of all, the ability to write
explored state spaces to disk will open up the possibility to postprocess and further analyse the state spaces. This could be done directly, or after application of bisimulation reduction on the GPU~\cite{wijs.tacas15}.

Second of all, we will work on making the tool more user friendly. Currently, providing an input model is quite cumbersome, since \GPUexplore requires a user to express system behaviour in the low level description formalism of networks of LTSs. Specifying systems would be much more convenient if a higher-level modelling language would be supported. We will investigate which modelling languages would be most suitable for integration in the current tool.

Finally, we will also consider the application of \GPUexplore to conduct computations similar to model checking, such as performance analysis~\cite{wijs.fokkink:chit.mcrl}. This requires to incorporate time into the input models, for instance by including special actions to represent the passage of time~\cite{wijs:drt}.
\bibliographystyle{eptcs}
\bibliography{paper}
\end{document}